\documentclass[prc,aps,superscriptaddress,showpacs,nofootinbib,twocolumn]{revtex4-1}
\usepackage{amssymb}
\usepackage[dvips]{graphicx}
\usepackage[english]{babel}
\usepackage{indentfirst}
\usepackage{amsxtra}
\usepackage{amsmath}
\usepackage{multirow}
\usepackage[mathcal]{eucal}
\usepackage{xspace}
\usepackage{hyperref}
\usepackage[xcolor=svgnames]{changes}

\newcommand{\br}{{\bf r}}

\begin{document}
\title {\bf Subtraction method in the second random--phase approximation: first
applications with a Skyrme energy functional}

\author{D. Gambacurta}
\affiliation{Dipartimento di Fisica e Astronomia and INFN, Via Santa Sofia 64, I-95123, Catania, Italy}

\author{M. Grasso}
\affiliation{Institut de Physique Nucl\'eaire, IN2P3-CNRS, Universit\'e Paris-Sud, 
F-91406 Orsay Cedex, France}

\author{J. Engel}
\affiliation{Department of Physics and Astronomy, University of North Carolina, Chapel Hill, North Carolina 27516-3255, USA}

\begin{abstract} 
We make use of a subtraction procedure, introduced to overcome double--counting
problems in beyond--mean--field theories, in the second
random--phase--approximation (SRPA) for the first time.  This procedure
guarantees the stability of SRPA (so that all excitation energies are real).
We show that the method fits perfectly into nuclear density--functional
theory.  We illustrate applications to the monopole and quadrupole response and
to low--lying $0^+$ and $2^+$ states in the nucleus $^{16}$O.  We show that the
subtraction procedure leads to: (i) results that are weakly cutoff dependent;
(ii) a considerable reduction of the SRPA downwards shift with respect to the
random--phase approximation (RPA) spectra (systematically found in all previous
applications).  This implementation of the SRPA model will allow
a reliable analysis of the effects of
2 particle--2 hole configurations ($2p2h$) on the excitation spectra of
medium--mass and heavy nuclei. 
\end{abstract} 

\vskip 0.5cm  \maketitle 
% 

%-----------------------------------------------------------------------
\section{Introduction}

Energy--density functional (EDF) theories have evolved over the years with the
formulation and application of sophisticated methods that go beyond mean--field
theory.  One example is the SRPA model, formulated long ago
\cite{providencia,Yannouleas} but applied in full only very recently because of
the extreme numerical effort required.  The last few
years have seen large--scale SRPA calculations done without approximations in
the matrices and with high--energy cutoffs
\cite{papa1,papa2,gamba1,gamba2,gamba3,gamba4}. Performing such calculations
has allowed us to identify some specific features of the SRPA model, that could
not be seen in previous strongly truncated and simplified calculations.
Unexpectedly, the SRPA spectrum is systematically lowered by several MeV with
respect to that obtained in the ordinary RPA.  The origin of this strong shift
was unclear until recently.

One would think that the SRPA, which adds $2p2h$ states to the 1 particle--1
hole ($1p1h$) states of the RPA, should not greatly modify excitations in which
$1p1h$ configurations represent the dominant contribution. The improvements
provided by the SRPA should be the following: 1) For cases in which the $1p1h$
states are the most important configurations and the RPA properly describes the
location of the main peaks of the response function, the SRPA should not change
those locations much. In addition, the coupling to $2p2h$ states should provide
some spreading and improve the calculated widths, which in the RPA are far too
small.  2) For cases in which multiparticle--multihole configurations are
essential to describe the excitation, the SRPA model should yield not only
spreading widths, but also a substantially different description of the overall
strength distribution than does the RPA.  Only recently it has started to
become clear that the unexpected shift in SRPA energies is intimately related
to the implicit inclusion of correlations in ground states;
properly accounting for those correlations
would guarantee a stable ground state and thus a consistent response function.

In Refs.\ \cite{gamba1,gamba2,gamba3,gamba4}, SRPA calculations were performed
with density--dependent effective interactions --- the Skyrme
\cite{skyrme1,skyrme2,vautherin} and the Gogny \cite{rifeGogny1,rifeGogny2}
interactions --- that derive from EDFs and are used extensively in medium--mass
and heavy nuclei.  Kohn--Sham--based density--functional theory (see next
section) requires that EDF parameters be adjusted through mean--field
calculations to reproduce properties of nuclear matter as well as masses and
radii of some selected nuclei. However, mean--field calculations, even with
density--dependent interactions, fail to reproduce several aspects of nuclear
phenomena.  Higher--order corrections (and correlations) may be explicitly
introduced by going beyond the mean--field approximation.  One must then depart
from strict density--functional theory and treat the interaction as an explicit
Hamiltonian.  Such beyond--mean--field calculations may lead to several
problems associated with mixing EDFs (which generate effective interactions
with non--antisymmetric matrix elements) and genuine Hamiltonian many--body
theory.  Moreover, when higher--order corrections are explicitly introduced,
the parameters of the interaction should be readjusted to avoid problems
of double counting \cite{moghrabi}.  In the particular
framework of extended RPA theories, such as SRPA, this would imply that
different interactions are used for the description of the ground state and of
excited states, with a lack of self--consistency leading to possible admixing
of spurious states in the physical spectrum. 

Some years ago \cite{tsela2007}, Tselyaev proposed another procedure to avoid
double--counting, called the ``subtraction'' method. It has been applied
thus far mainly to models that include particle--vibration coupling,
specifically both the non--relativistic \cite{nonrela1,nonrela2,nonrela3} and
relativistic \cite{rela1,rela2,rela3} versions of the time--blocking
approximation.  More recently \cite{tsela2013}, Tselyaev formally demonstrated
that this procedure guarantees the validity of the usual stability condition in
extensions of the RPA.  The stability of the RPA is related to the Thouless
theorem \cite{thouless}, which states that the Slater determinant on which the
RPA is based must be a minimum of the Hartree--Fock (HF) energy to guarantee real RPA
eigenvalues. Ref.\ \cite{tsela2013} showed that this theorem does not hold in
general for extended versions of the RPA, such as the SRPA, and that the
subtraction procedure reinstates it.  

For the SRPA with a real density--independent Hamiltonian, the problem of
stability was addressed in a different way by Papakonstantinou
\cite{papa2014}.  The author showed that: (i) the Thouless theorem can be
extended to the SRPA model if an explicitly correlated ground state is used;
(ii) in spite of the stability problem, the energy--weighted sum rule (EWSR)
\cite{thouless,ringshuck} is satisfied within the SRPA model, confirming a
previous study by Yannouleas \cite{Yannouleas}.  Indeed, an extension of SRPA
with a correlated ground state was applied to metallic clusters in the work
reported in Ref.\ \cite{gambacatara}.  The standard SRPA with a HF ground state
produced a significant shift of strength towards lower energies with respect to
the RPA strength function. The use of a correlated ground state, however,
pushed the strength back towards larger energies, closer to the RPA (and
experimental) results.  In this kind of calculation, the shift is clearly due
to a consistent ground state, of which an account of correlations and stability
are both clearly important features. 

As already mentioned, however, explicit correlations (for example, a correlated
ground state in SRPA) may be problematic in an EDF--based calculation. The
ground state of an EDF--based model already has an energy and density designed
to be as close as possible to the exact ones.  The Tselyaev subtraction
procedure, by contrast, fits beautifully into density functional theory, as we
show in the next section, and is much less costly numerically.  It has in
addition the great advantage of simultaneously
removing double--counting problems and instabilities.  This is
the procedure adopted in the present work. 

In subsequent sections, we apply the subtraction procedure within the SRPA for
the first time, to the monopole and the quadrupole giant resonances of the
nucleus $^{16}$O, as well as to its low--lying $0^+$ and $2^+$ states.  We will
compare the results with those of the ordinary RPA and with experiment.  We
perform the calculations with the Skyrme functional SGII \cite{sgii}, both in
the mean--field equations and in the effective residual interaction.  We omit
Coulomb and spin--orbit contributions in the residual interaction but include
rearrangement terms following Ref.\  \cite{gamba2}.  

The article is organized as follows: Section II discusses the subtraction
method in the context of density--functional theory.
Section III presents our formalism and discusses a convenient ``diagonal''
approximation.  Section IV illustrates the results obtained with the 
subtracted SRPA, and assesses the accuracy of the diagonal approximation.
Section V compares subtracted SRPA response functions with those of the
ordinary RPA and with experiment. Section VI presents conclusions.

\section{Subtraction Method and Density--Functional Theory}

It is difficult to say exactly what is meant by a density--dependent
interaction, particularly when it does not have antisymmetric two--body matrix
elements.  In practice such an interaction is always used to construct an
expression for the energy, which is then varied to obtain mean--field
equations.  It turns out to be natural not to worry about the interaction and
to work instead with the energy as a functional of various densities --- the
ordinary number density, the spin--density, etc.  The condensed--matter and
atomic--physics communities have seized on this idea, the elaboration of which
is known as density functional theory \cite{dft}.

The foundation of the theory consists of the Hohenberg--Kohn theorem
\cite{hohenbergkohn} and the Kohn--Sham procedure \cite{kohnsham}.  These
building blocks have to be modified slightly for nuclear physics so that they
work with densities that are defined with respect to the nuclear center of
mass, but they survive the modification intact \cite{engel}.  To simplify
matters here, we pretend that only the intrinsic one--body density $\rho(\br)$
is relevant; that is, we neglect other densities and currents on which the
functional typically depends.  Density functional theory then works with the
object $E[\rho]$, the meaning of which is the smallest expectation value of the
underlying nuclear Hamiltonian produced by states that yield density $\rho$.
Thus, one finds the system's ground--state energy and density by minimizing
$E[\rho]$.  The Hohenberg--Kohn theorem states that the energy--density
functional $E[\rho]$ is universal: in the presence of an additional local
operator $\lambda Q(\br)$, with $\lambda$ an arbitrary constant, $E[\rho]$ is
modified in a simple way, 
\begin{equation}
\label{eq:univ}
E[\rho] \longrightarrow E_\lambda[\rho] = E[\rho] + \lambda \! \int \! d\br \, Q(\br) \rho(\br). 
\end{equation}
The Kohn--Sham procedure guarantees that any energy--density functional can be
written in terms of orbitals $\varphi_i(\br)$, as if the system consisted of
noninteracting particles in a density--dependent external potential, and that
the ground state energy and density can be found by solving the Hartree--like
equations for the orbitals that come from minimizing the functional with
respect to the orbital wave functions.  Skyrme and Gogny energy--density
functionals for a given nucleus make perfect sense when interpreted as
approximations to the Kohn--Sham functional.  

Now suppose that the multiplier $\lambda$ is small.  Then, system's density
changes from the unperturbed ground--state density $\rho_0$ to a new one
$\rho_\lambda$, given by
\begin{equation}
\label{eq:pert-dens}
\rho_\lambda = \rho_0+ \lambda \int \! d\br \, R(\omega=0,\br,\br') Q(\br), 
\end{equation}
where $R(\omega=0)$ is the static response function for the underlying
Hamiltonian.  Because the energy functional is universal, it reproduces
$\rho_\lambda$ exactly when modified as in Eq.\ \ref{eq:univ}.  And in the
Kohn-Sham approach, the functional produces a mean--field effective Hamiltonian
(that nonetheless reproduces exact energies and densities), so that the
response function $R(\omega=0)$ is given by the RPA, which is the
small--amplitude limit of time--dependent mean--field theory \cite{ringshuck}. In other words
\begin{equation}
\label{eq:resp-zero}
R(\omega=0) = R^{RPA}_{KS} \,,
\end{equation}
where $R^{RPA}_{KS}$ is the RPA response associated with the Kohn--Sham
representation of the functional and the corresponding ground--state Slater
determinant.  Thus, to the extent that the Skyrme functional is the exact
Kohn--Sham functional, Skyrme--RPA produces the exact zero--frequency response
function, and any modification of the response must vanish in the static
limit.  This is what is meant by ``avoiding double counting,'' a fact that was
noted by Tselyaev in Ref.\ \cite{tsela2013}.

More generally, one can show through a time--dependent version of the
Hohenberg--Kohn theorem (known as the Runge--Gross theorem \cite{rungegross}) and
a time--dependent Kohn--Sham procedure that the full response function at any
frequency obeys 
\begin{align}
\label{eq:resp-eq}
R(\omega,&\br,\br') = R^0_{KS}(\omega,\br,\br')\\
&+ \int \! d\br_1 d\br_2 \,
R^0_{KS}(\omega,\br, \br_1) V(\omega,\br_1,\br_2) R(\omega,\br_2,\br') \,,
\nonumber
\end{align}
where $R^0_{KS}$ is the bare Kohn--Sham (mean--field) response and $V(\omega)$ is
a frequency--dependent effective interaction obtained from the time--dependent
energy--density functional $\mathcal{E}[\rho(t),t]$.  The approximation
\begin{equation}
\label{eq:adiab-approx}
V(\omega,\br_1,\br_2) \longrightarrow \frac{\delta^2
E[\rho]}{\delta \rho(\br_1) \delta \rho (\br_2)}\bigg|_{\rho_0} \equiv
V^{RPA}(\br_1,\br_2) \,,
\end{equation}
implies that the solution $R$ is just $R^{RPA}_{KS}$, which does not depend on
$\omega$. The approximation is equivalent to assuming that
$\mathcal{E}[\rho(t),t] = E[\rho(t)]$, that is, that the time--dependent energy is
just the ground--state functional evaluated at the time--dependent density.
Making that approximation is known as the adiabatic limit.
To go beyond that limit, one must introduce an $\omega$ dependence into the
effective two--body interaction $V$.  That, as we will see shortly, is precisely
what the SRPA does.  But since $R^{RPA}_{KS}$ is correct (as correct as the
Skyrme functional, anyway) in the adiabatic limit, we must modify the SRPA so
that it gives the RPA response at $\omega=0$.  

There are many ways one might modify the SRPA $V(\omega)$ so as to obtain the
RPA response in the adiabatic limit. One could, for example, simply multiply
$V^{SRPA}(\omega)$ by a function that is unity at large
$\omega$ and falls to $V^{RPA}/V^{SRPA}(0)$ as $\omega$ goes to zero.  But the
response function has other constraints as well; in particular the quantity
$\text{Im}[\int d\br d\br' Q(\br) R(\omega,\br,\br') Q(\br')]$ must have real
and positive residues at poles on the positive real axis, that is, it must
produce a genuine strength function (the stability condition).  Although there
may be more than one way to ensure this, a particularly simple way is the
subtraction method.  If we define the frequency/energy--dependent difference
between the SRPA and RPA effective interactions by $U(\omega)$,
\begin{equation}
\label{eq:U}
U(\omega) \equiv V^{SRPA}(\omega)-V^{RPA}(\omega) \,,
\end{equation}
the subtraction procedure amounts to the replacement 
\begin{equation}
\label{eq:subtraction-dft}
V^{SRPA}(\omega) \longrightarrow V^{SRPA}(\omega) - U(0) \,,
\end{equation}
which guarantees, as required, that $V^{SRPA}(0)=V^{RPA}(0)$ after the
substitution and thus that this ``subtracted SRPA'' reduces to the RPA in the
zero--frequency limit.

In the next section we describe how the method works in the matrix version of
the SRPA.
 
\section{Subtraction method in SRPA}

Here we work with the matrix formulation of the SRPA.  Details about
the associated formalism appear, for instance, in Ref.\ \cite{Yannouleas}.
Details about the subtraction method and its specific application to extended
RPA models can be found in Refs. \cite{tsela2007,tsela2013}.  

The SRPA equations can be written in the following form:  
\begin{equation}\label{eq_srpa}
\left(\begin{array}{cc}
  \mathcal{A} & \mathcal{B} \\
  -\mathcal{B}^{*} & -\mathcal{A}^{*} \\
\end{array}\right)
\left(%
\begin{array}{c}
  \mathcal{X}^{\nu} \\
  \mathcal{Y}^{\nu} \\
\end{array}%
\right)=\omega_{\nu}
\left(%
\begin{array}{c}
  \mathcal{X}^{\nu} \\
  \mathcal{Y}^{\nu} \\
\end{array}%
\right),
\end{equation}
where
\begin{equation}
\mathcal{A}=\left(\begin{array}{cc}
  A_{11'} & A_{12} \\
  A_{21} & A_{22'} \\
\end{array}\right), \quad
% % \end{displaymath}
% \begin{displaymath}
\mathcal{B}=\left(\begin{array}{cc}
  B_{11'} & B_{12} \\
  B_{21} & 0\\ %%B_{22'} \\
\end{array}\right),
\end{equation}
\begin{equation}
\label{xandy}
\mathcal{X}^{\nu}=\left(\begin{array}{cc}
  X_{1}^{\nu} \\
   X_{2}^{\nu} \\
\end{array}\right),
~~~~\mathcal{Y}^{\nu}=\left(\begin{array}{cc}
  Y_{1}^{\nu} \\
   Y_{2}^{\nu} \\
\end{array}\right).
\end{equation}

The indices 1 and 2 denote $1p1h$ and $2p2h$ configurations, respectively.
Thus, the $11^{\prime}$ block in the matrices $\mathcal{A}$ and $\mathcal{B}$
corresponds to the standard RPA $A$ and $B$ matrices.  The 12 and 21 blocks
contain the $1p1h-2p2h$ coupling, and the $22^{\prime}$ block (the $B$ part of
which vanishes) contains the $2p2h$ part of the matrix. 

It is straightforward to show that the SRPA equations may be written as
RPA--like equations with energy--dependent RPA matrices
$A_{11^{\prime}}(\omega)$ and $B_{11^{\prime}}(\omega)$, 
\begin{widetext}
\begin{eqnarray}
\label{arpa}
A_{11^{\prime}} (\omega) = A_{11^{\prime}}+\sum_{2,2^{\prime}} A_{12} (\omega + i \eta - A_{22^{\prime}})^{-1} 
A_{2^{\prime}1^{\prime}} - 
\sum_{2,2^{\prime}} B_{12} (\omega + i \eta + A_{22^{\prime}})^{-1} 
B_{2^{\prime}1^{\prime}} \,,\\
B_{11^{\prime}} (\omega) = B_{11^{\prime}} + \sum_{2,2^{\prime}} A_{12} (\omega + i \eta - A_{22^{\prime}})^{-1} 
B_{2^{\prime}1^{\prime}} -  
\sum_{2,2^{\prime}} B_{12} (\omega + i \eta + A_{22^{\prime}})^{-1} 
A_{2^{\prime}1^{\prime}}\,. \nonumber
\label{brpa} 
\end{eqnarray}
These are just the analogs of the $\omega$--dependent interaction $V(\omega)$ from
Sec. II in the more general particle--hole basis.  Without so--called
rearrangement terms, $B_{12}$ and $B_{21}$ would vanish, there would be no
correction to $B_{11^{\prime}}$, which would simply become the corresponding
RPA matrix (no energy dependence) and the last term in Eq. (\ref{arpa}) would
be zero.  We include rearrangement terms here, however. 

Let us denote by $E_{11^{\prime}}(\omega)$ and $F_{11^{\prime}}(\omega)$ the
energy--dependent corrections to $A_{11^\prime}$, and $B_{11^\prime}$:

\begin{eqnarray}
E_{11^{\prime}} (\omega) =  \sum_{2,2^{\prime}} A_{12} (\omega + i \eta - A_{22^{\prime}})^{-1} 
A_{2^{\prime}1^{\prime}}  - 
\sum_{2,2^{\prime}} B_{12} (\omega + i \eta + A_{22^{\prime}})^{-1} 
B_{2^{\prime}1^{\prime}}\,, \\
F_{11^{\prime}} (\omega) = \sum_{2,2^{\prime}} A_{12} (\omega + i \eta - A_{22^{\prime}})^{-1} 
B_{2^{\prime}1^{\prime}} - 
\sum_{2,2^{\prime}} B_{12} (\omega + i \eta + A_{22^{\prime}})^{-1} 
A_{2^{\prime}1^{\prime}}\,. \nonumber
\end{eqnarray}

\end{widetext}
The subtraction procedure, in this matrix context, corrects the RPA--like
matrices by subtracting from $A_{11^{\prime}}(\omega)$ and
$B_{11^{\prime}}(\omega)$ the static parts $E_{11^{\prime}}(0)$ and
$F_{11^{\prime}}(0)$, respectively (the analogs of $U$ in Sec. II):
\begin{equation}
A^S_{11^{\prime}} (\omega)= A_{11^{\prime}} (\omega) - E_{11^{\prime}}(0),
\label{sub1} 
\end{equation}
\begin{equation}
B^S_{11^{\prime}} (\omega)= B_{11^{\prime}} (\omega) - F_{11^{\prime}}(0). 
\label{sub2} 
\end{equation}
$A^S_{11^{\prime}} (\omega)$ and $B^S_{11^{\prime}} (\omega)$ are then
substituted for $A_{11^{\prime}} (\omega) $ and $B_{11^{\prime}} (\omega)$ in
the energy--dependent RPA--like equations.  One can then return to
energy--independent equations with larger matrices, obtaining
\begin{widetext}
\begin{eqnarray}
\label{eq:absf}
\mathcal{A}^S_F=\left(\begin{array}{cc}
  A_{11'}+ \sum_{2,2'} A_{12} (A_{22'})^{-1}A_{2'1'} + \sum_{2,2'} B_{12} (A_{22'})^{-1} B_{2'1'} & A_{12} \\
  & \\
  A_{21} & A_{22'} \\
\end{array}\right)\,, \\[.3cm]
\mathcal{B}^S_F=\left(\begin{array}{cc}
  B_{11'} + \sum_{2,2'} A_{12} (A_{22'})^{-1} B_{2'1'} + \sum_{2,2'} B_{12} (A_{22'})^{-1} A_{2'1'}  & B_{12} \\
  &  \\
  B_{21} & 0\\ %B_{22'} \\
\end{array}\right)\,. \nonumber
\end{eqnarray}
\end{widetext}
The subscript $F$ stands for ``full'' here.  We will contrast the full calculation later with a
``diagonal approximation''. We note that the restoration of the RPA in the
$\omega=0$ limit implies that the static polarizability, related to the inverse
energy--weighted moment $m_{-1}$, is the same in the subtracted SRPA as in the
RPA.  Because $m_{-1}$ is generally larger in the standard SRPA than in the
RPA, this fact already suggests that subtraction, which restores the RPA
$m_{-1}$, must shift energy upwards with respect to the ordinary SRPA.  

The matrix $A_{22'}$ has to be inverted to construct the matrices in Eq.\
\ref{eq:absf}.  If the energy cutoff is very large, the dimensions of the
matrix become even more so and the inversion becomes computationally costly.
We therefore will sometimes use a diagonal approximation in which off--diagonal
terms of the matrix $A_{22^\prime}$ are discarded only in the correction
terms.  The subtracted SRPA matrices, which we call $\mathcal{A}^S_{D}$
and $\mathcal{B}^S_{D}$, then are
\begin{widetext}
\begin{eqnarray}
\label{diagoo1}
\mathcal{A}^S_D=\left(\begin{array}{cc}
  A_{11'}+ \sum_2 A_{12} (A^{diag}_{22})^{-1}A_{21'} + \sum_2 B_{12} (A^{diag}_{22})^{-1} B_{21'} & A_{12} \\
 & \\
  A_{21} & A_{22'} \\
\end{array}\right)\,,\\[.2cm] 
\mathcal{B}^S_D=\left(\begin{array}{cc}
  B_{11'} + \sum_2 A_{12} (A^{diag}_{22})^{-1} B_{21'} + \sum_2 B_{12}
  (A^{diag}_{22})^{-1} A_{21'}  & B_{12} \\
  & \\
  B_{21} & 0\\ %B_{22'} \\
\end{array}\right)\,. \nonumber 
\end{eqnarray}
\end{widetext}
With this approximation the computational effort is considerably reduced.  When
Eq. (\ref{diagoo1}) is used, one can either keep only the unperturbed energies
or include also the residual interaction in $A^{diag}_{22}$ to compute the
correction terms. We have verified that the two choices lead to very similar
results. In what follows, we show the results obtained by neglecting the
residual interaction.  

%\textcolor{red}{In next sections, we compare the full SRPA scheme with the full subtraction procedure [Eqs.
%(10) and (11)] will be denoted with $F$, whereas the SRPA scheme with the
%approximation of Eqs. (12) and (13) will be called $DCorr$.}

\section{Subtracted spectra and accuracy of the diagonal
approximation}

\begin{figure}
\includegraphics[scale=0.35]{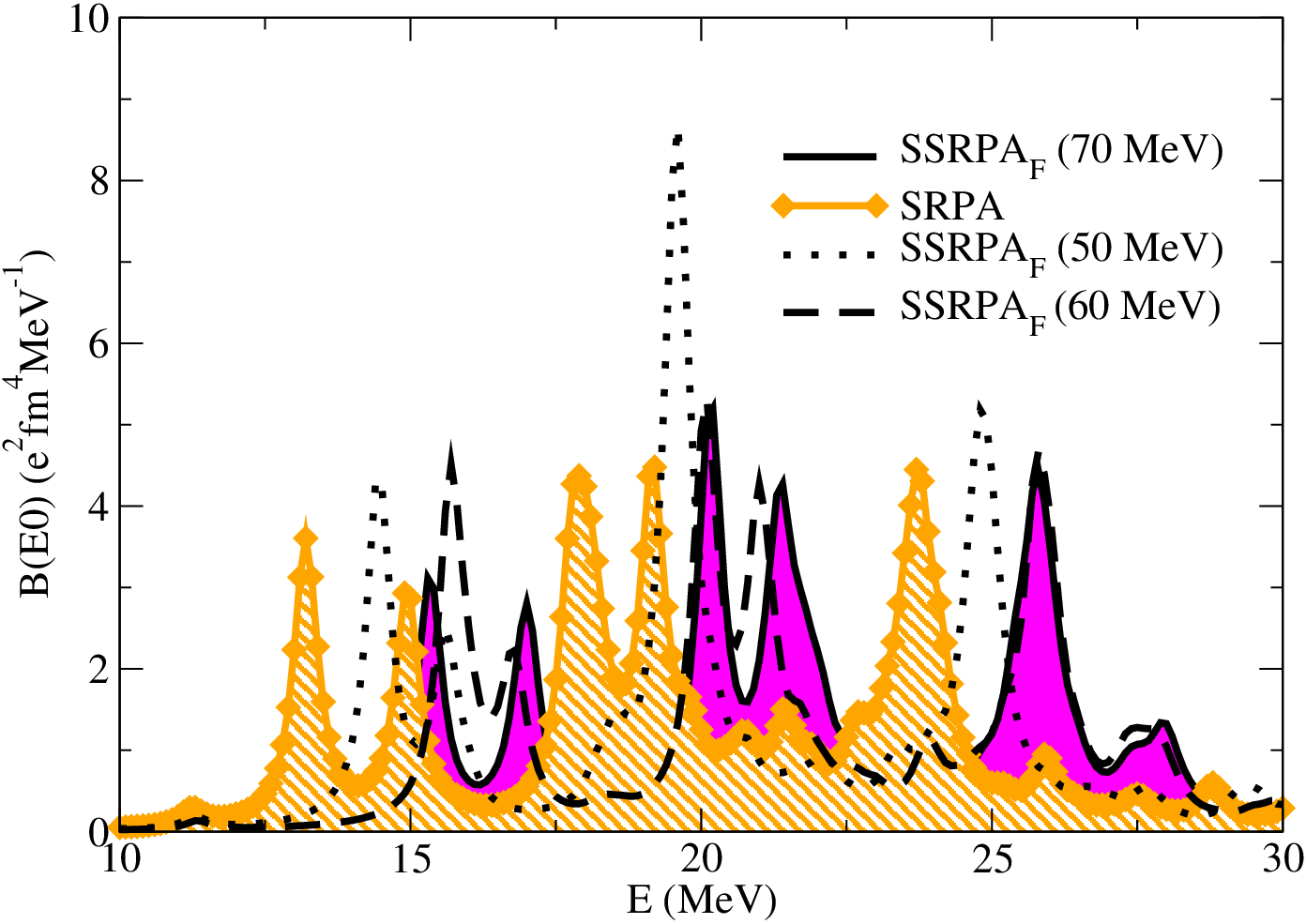}
\caption{(Color online) Isoscalar monopole response for the nucleus $^{16}$O,
calculated in the standard SRPA (orange diamonds and orange area), and with
the SSRPA$_F$, with a cutoff for the correction
terms at 50 (black dotted line), 60 (black
dashed line), and 70 (black solid line and magenta area) MeV.}
\label{monopole1}
\end{figure}

\begin{figure}
\includegraphics[scale=0.35]{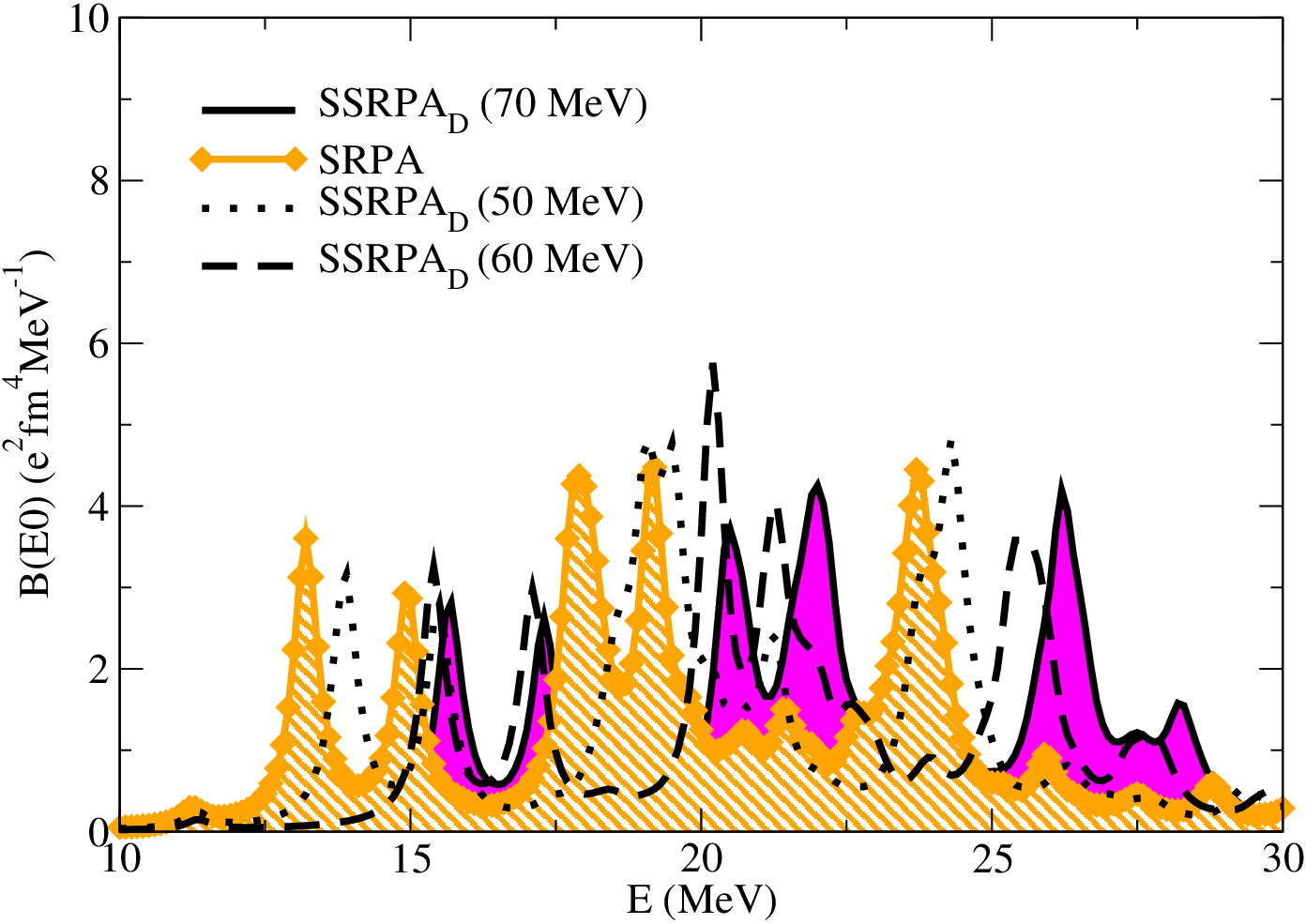}
\caption{(Color online) Same as in Fig.\ \ref{monopole1}, but in the diagonal
approximation SSRPA$_D$.
\label{monopole2}}
\end{figure}

Let us first show the effect of the subtraction method on the SRPA results,
for illustration in the isoscalar monopole and quadrupole channels of
$^{16}$O.  We perform HF--RPA calculations with a cutoff on the $1p1h$
configurations at 100 MeV. For the $2p2h$ space in the SRPA calculations, we
take the cutoff to be at 70 MeV and 50 MeV for the monopole and the quadrupole
cases, respectively.  Those values lead to about 5000 $2p2h$ configurations in
each of the two cases. This number is small enough so that we can still
fully invert the matrix $A_{22\prime}$ to perform the subtraction.  

We use the acronyms SSRPA$_F$ to denote the subtracted SRPA in the full scheme,
Eq. (\ref{eq:absf}), and SSRPA$_D$ to denote the subtracted SRPA with the
diagonal approximation in the correction terms, Eq.
(\ref{diagoo1}).  In all the figures that follow, we fold the calculated
response with a Lorentzian of width 0.5 MeV. 

Fig.\ \ref{monopole1} shows the isoscalar monopole strength distribution,
calculated with the unmodified SRPA and with the SSRPA$_F$,
using a cutoff in the correction term equal to 50, 60, and 70 MeV. In the last
of these cases, all the SRPA $2p2h$ configurations are included in the
correction.  The effect of the subtraction, as we expected, is to shift the
SRPA spectrum upwards, by amounts that increase with the cutoff in the
correction terms.  The important differences between the three
subtracted strength functions indicate that it is crucial to include all the
$2p2h$ states in the correction terms containing $(A_{22^\prime})^{-1}$ in Eq.\
(15).  The calculation must be coherent, that is, the $2p2h$ spaces used in the
original SRPA matrices and in the correction terms should be the same.  
 
\begin{figure}[b]
\includegraphics[scale=0.35]{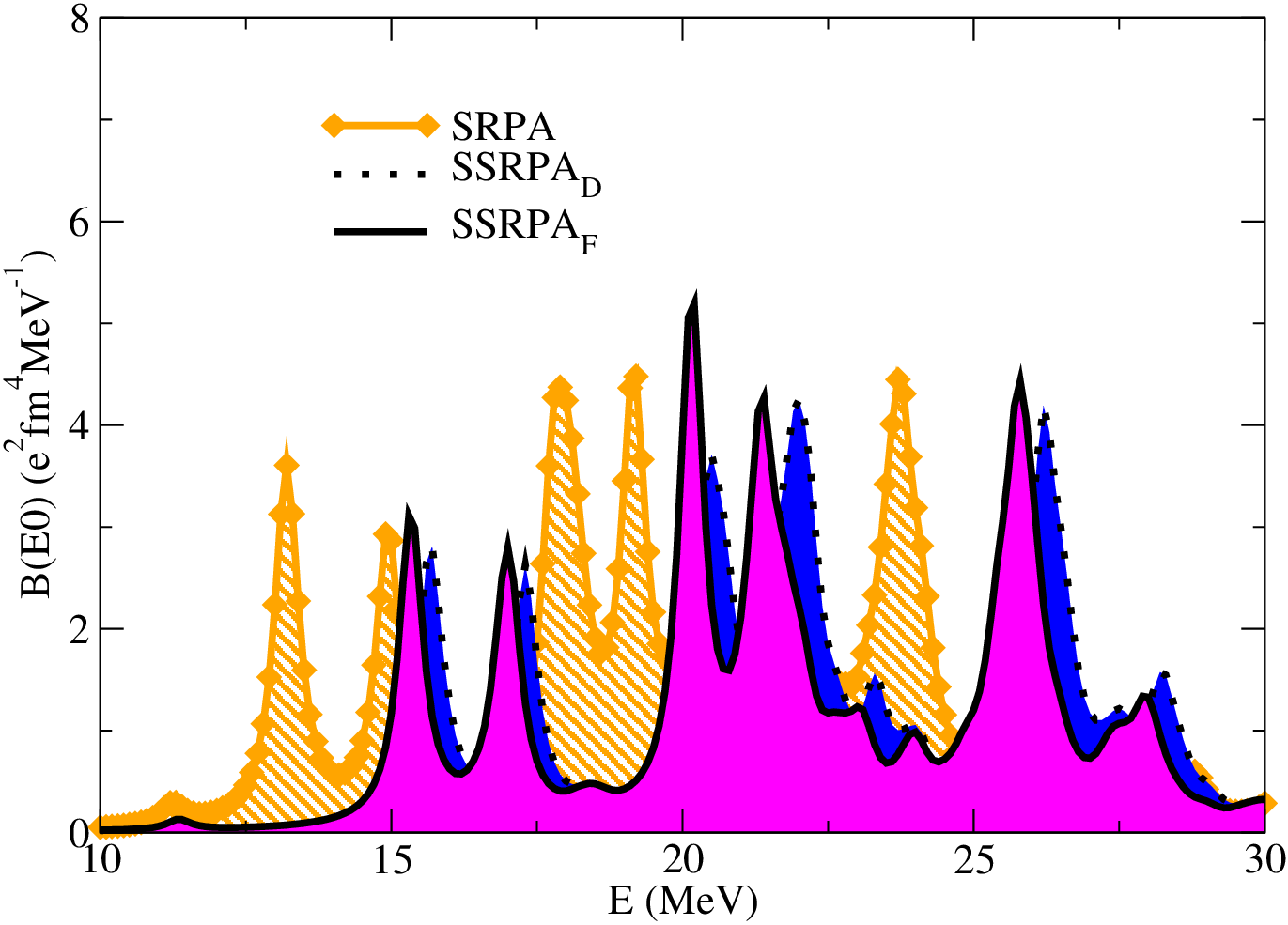}
\caption{(Color online) Isoscalar monopole response for the nucleus $^{16}$O,
calculated in the SRPA without subtraction (orange diamonds and orange area),
in the SSRPA$_F$ (black solid line and magenta area) and in
the SRPA$_D$ (black dotted line and blue area), with a cutoff in
the correction terms at 70 MeV. }
\label{monopole3}
\end{figure}
Fig.\ \ref{monopole2} shows the same results with the diagonal approximation
SSRPA$_D$ and Fig.\ \ref{monopole3} compares the full and diagonal subtracted
SRPA results with the 70--MeV cutoff in the correction terms.  We observe that
the SSRPA$_F$ and SSRPA$_D$ results are very similar, the difference being a
small systematic shift to larger excitation energies in the
SSRPA$_D$. 

\begin{figure}
\includegraphics[scale=0.35]{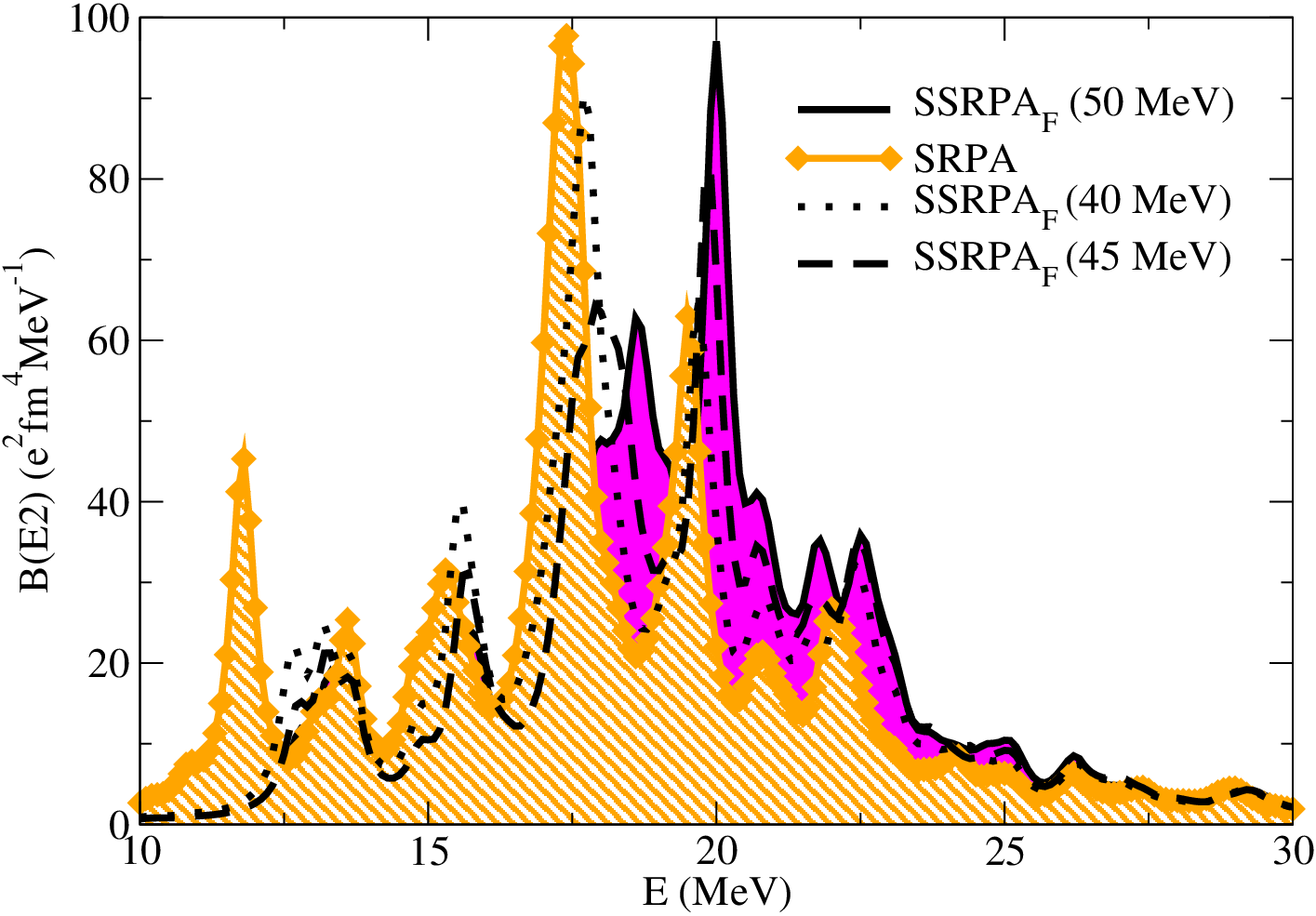}
\caption{(Color online) Same as in Fig.\ \ref{monopole1} but for the isoscalar
quadrupole response. The $2p2h$ cutoffs in the correction terms are now at 40,
45, and 50 MeV.}
\label{quadrupole1}
\end{figure}

\begin{figure}
\includegraphics[scale=0.35]{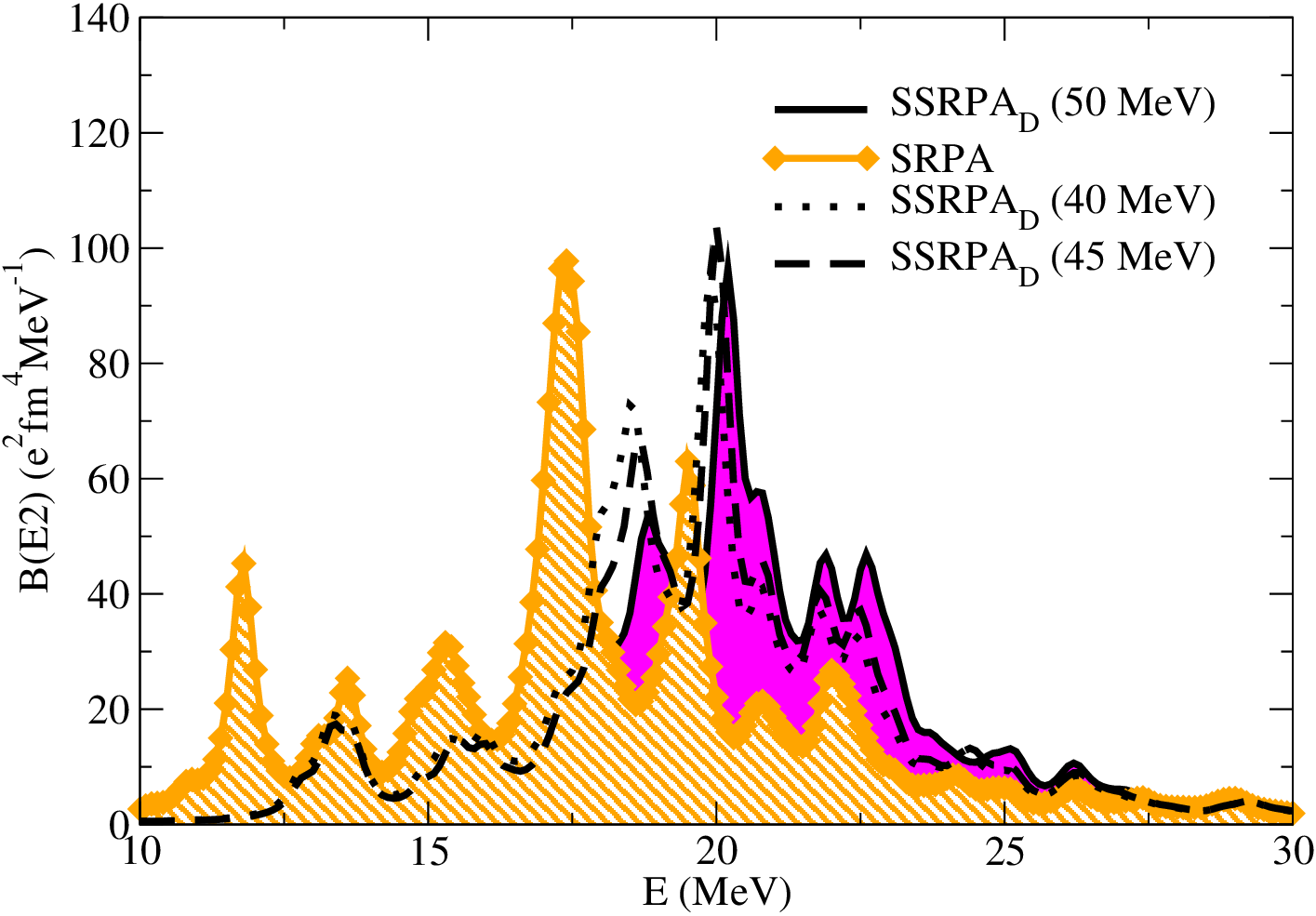}
\caption{(Color online) Same as in Fig.\ \ref{monopole2} but for the isoscalar
quadrupole response.}
\label{quadrupole2}
\end{figure}

\begin{figure}
\includegraphics[scale=0.35]{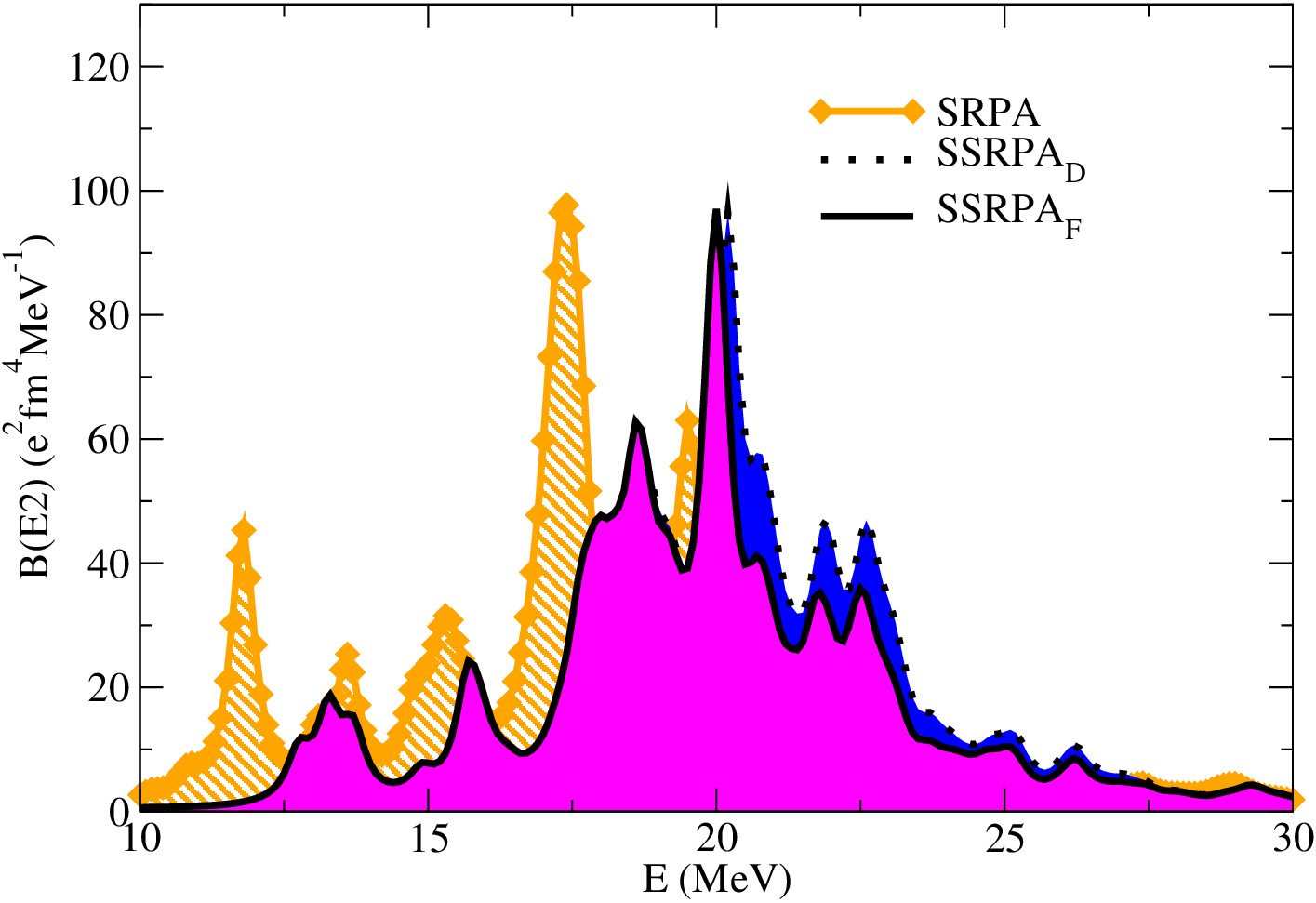}
\caption{(Color online) Same as in Fig.\ \ref{monopole3} but for the isoscalar
quadrupole response. The $2p2h$ cutoff for the correction terms is at 50 MeV.}
\label{quadrupole3}
\end{figure}

Figs.\ \ref{quadrupole1}, \ref{quadrupole2}, and \ref{quadrupole3} display the
same results as Figs. \ref{monopole1}, \ref{monopole2}, and \ref{monopole3},
respectively, but in the isoscalar quadrupole channel (where the $2p2h$ cutoff
is at 50 MeV in the SRPA calculation). The cutoffs in the correction terms are
at 40, 45, and 50 MeV. The remarks made about the first three figures apply
here as well.  In Fig.\ \ref{quadrupole3}, the same systematic shift between
the results of the full and diagonal subtracted calculations is visible. 

\begin{figure}
\includegraphics[scale=0.35]{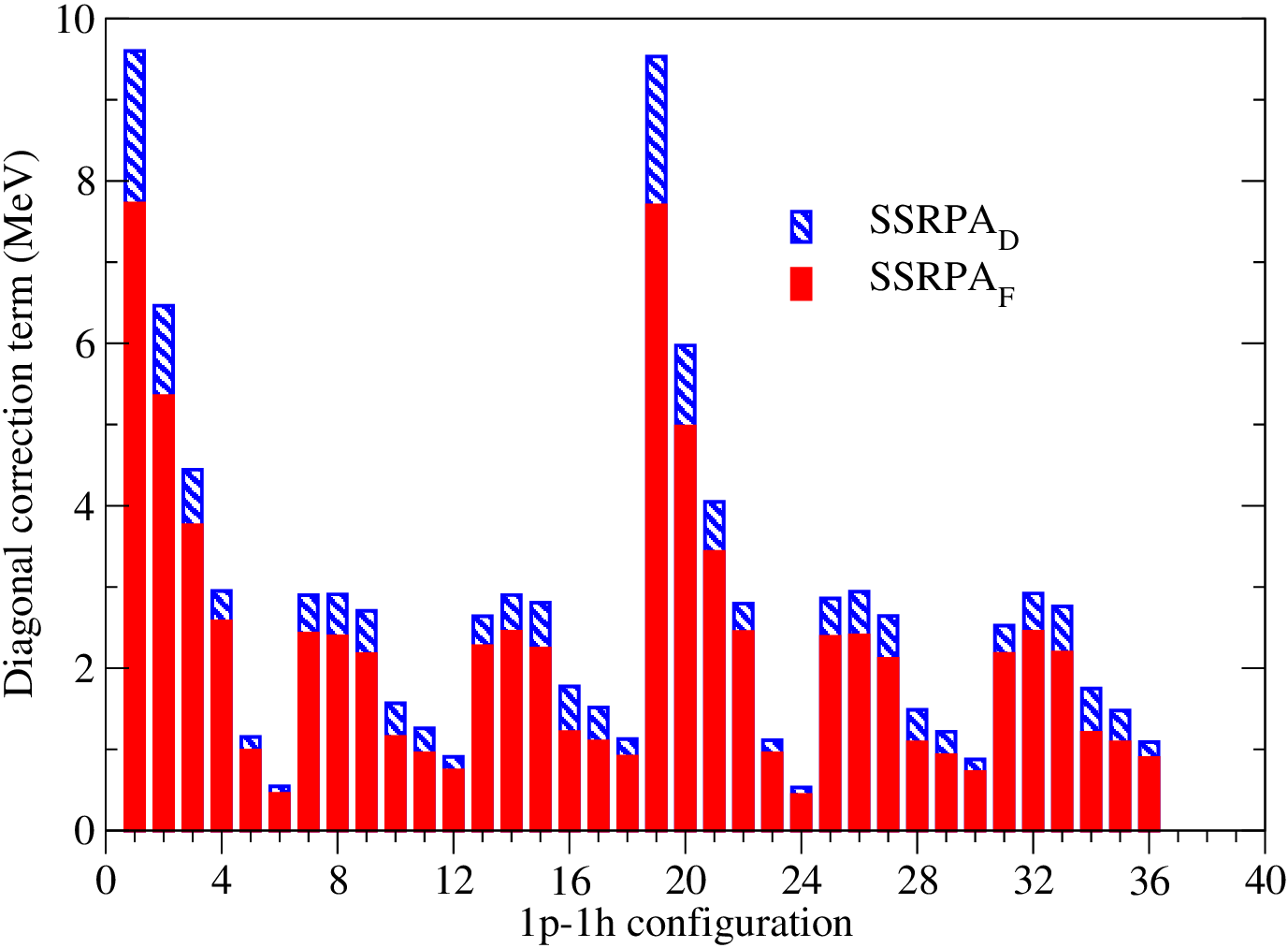}
\caption{(Color online) Diagonal part of correction term for each $1p1h$
configuration that contributes to the monopole strength, for the
SSRPA$_F$ (red full bars) and the SSRPA$_D$ (blue dashed bars),
with a cutoff at 70 MeV in the correction terms.} 
\label{diago1}
\end{figure}

\begin{figure}
\includegraphics[scale=0.35]{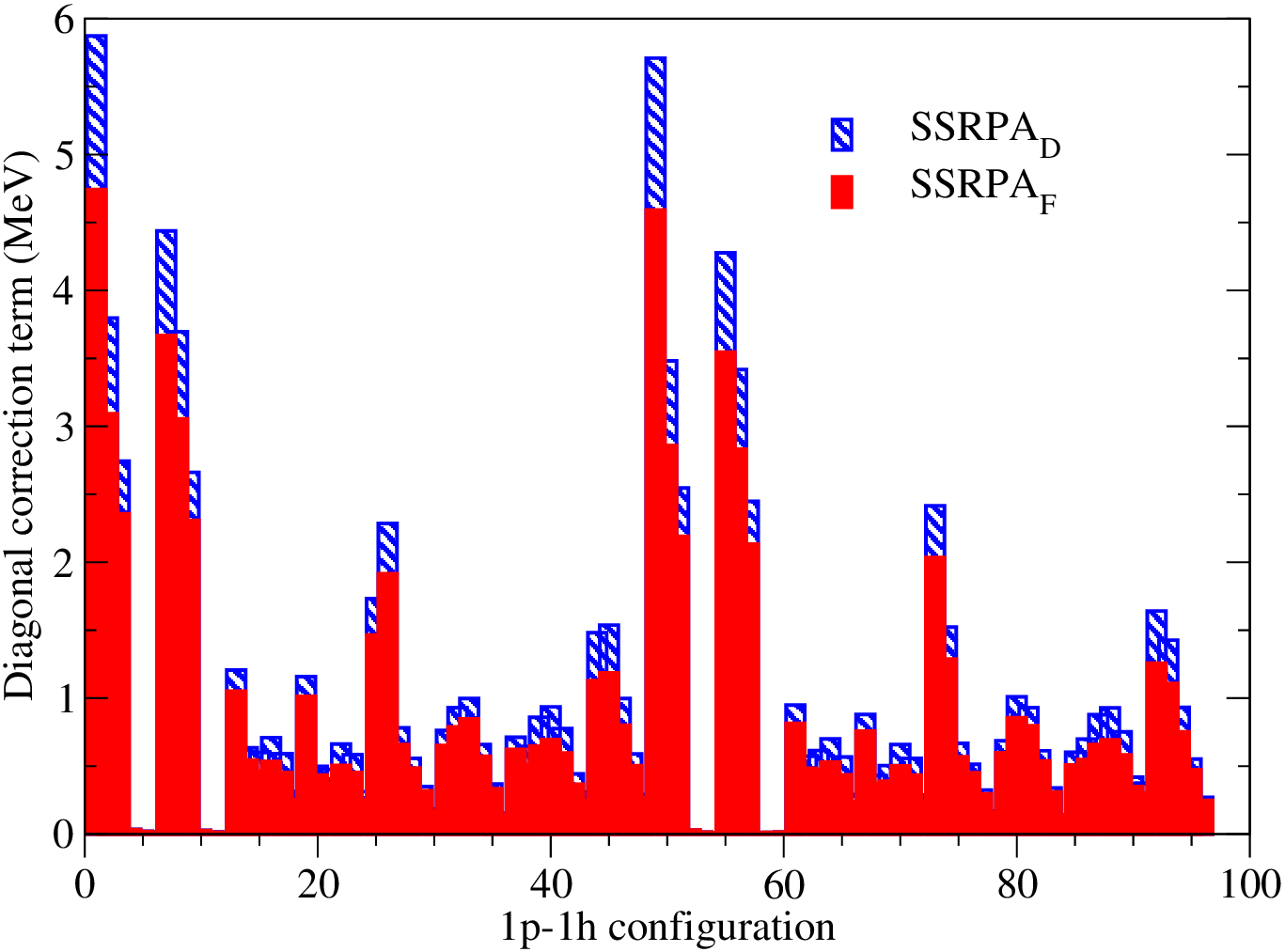}
\caption{(Color online) Same as in Fig.\ \ref{diago1}, but for the quadrupole
channel, with a cutoff at 50 MeV in the correction terms.}
\label{diago2}
\end{figure}

To better understand the extra shift produced by the SSRPA$_D$,
we plot in Figs.\ \ref{diago1} and \ref{diago2} the diagonal part of the
correction in the monopole and quadrupole channels, respectively, calculated
for each $1p1h$ state with the SSRPA$_F$ and SSRPA$_D$, with the
largest cutoff in each multipole.  The correction introduced
by the subtraction modifies the diagonal part of the
RPA $A$ matrix, and induces a shift in the $1p1h$ unperturbed excitation
energies.  The figures show that the diagonal correction term is always larger
in the SSRPA$_D$ than in the SSRPA$_F$.  This
leads to the extra shift of the spectrum found in the
SSRPA$_D$.  The difference, while systematic, is small. The result suggests
that the dominant effect of the correction comes from its diagonal part, which
modifies the unperturbed excitation energies. 

\begin{figure}
\includegraphics[scale=0.35]{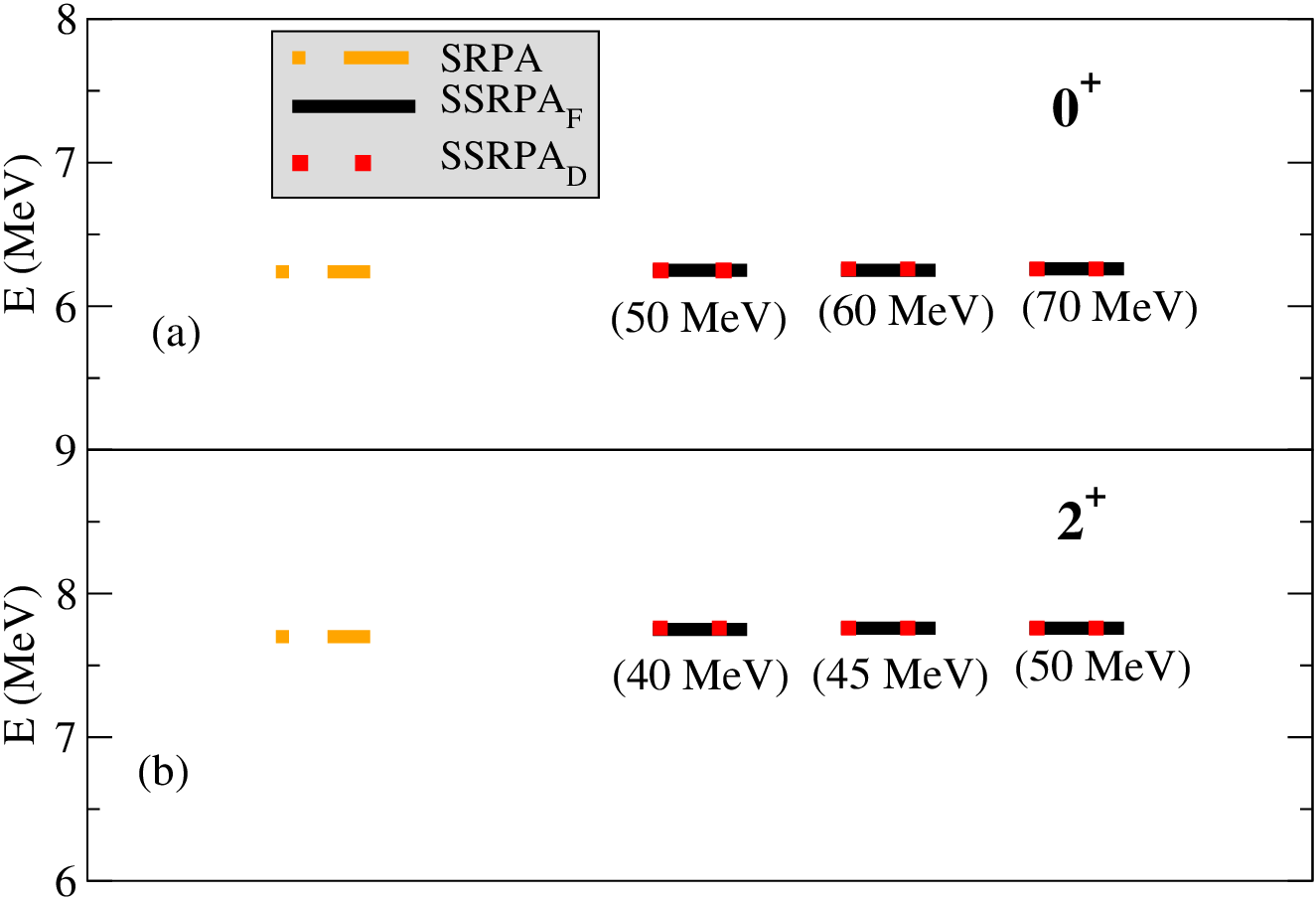}
\caption{(Color online) First 0$^+$ (a) and 2$^+$ (b) states calculated with
the standard SRPA, the SSRPA$_F$, and the
SSRPA$_D$, with different cutoffs in the correction terms (in parentheses).}
\label{low}
\end{figure}

We turn now to low--lying excited states.  Fig.\ \ref{low} shows the first
$0^+$ and $2^+$ states obtained in the unmodified SRPA and
in the SSRPA$_F$ and SSRPA$_D$, with
different $2p2h$ cutoffs in the correction terms.  Only states with a $B(E0)$
or $B(E2)$ larger than 10$^{-2}$ e$^2$fm$^4$ are shown.  Interestingly, the
effect of the correction is now different than
what we found for the giant resonances. Low--lying states are
not strongly modified by the subtraction method. This suggests that these
states have a dominant multiparticle--multihole nature, as a close examination
of the $\mathcal{X}$'s and $\mathcal{Y}$'s from Eq.\ (\ref{xandy}) verifies.
They cannot be affected by the subtraction because it acts only on the $1p1h$
sector of the SRPA matrix.

\begin{figure}
\includegraphics[scale=0.35]{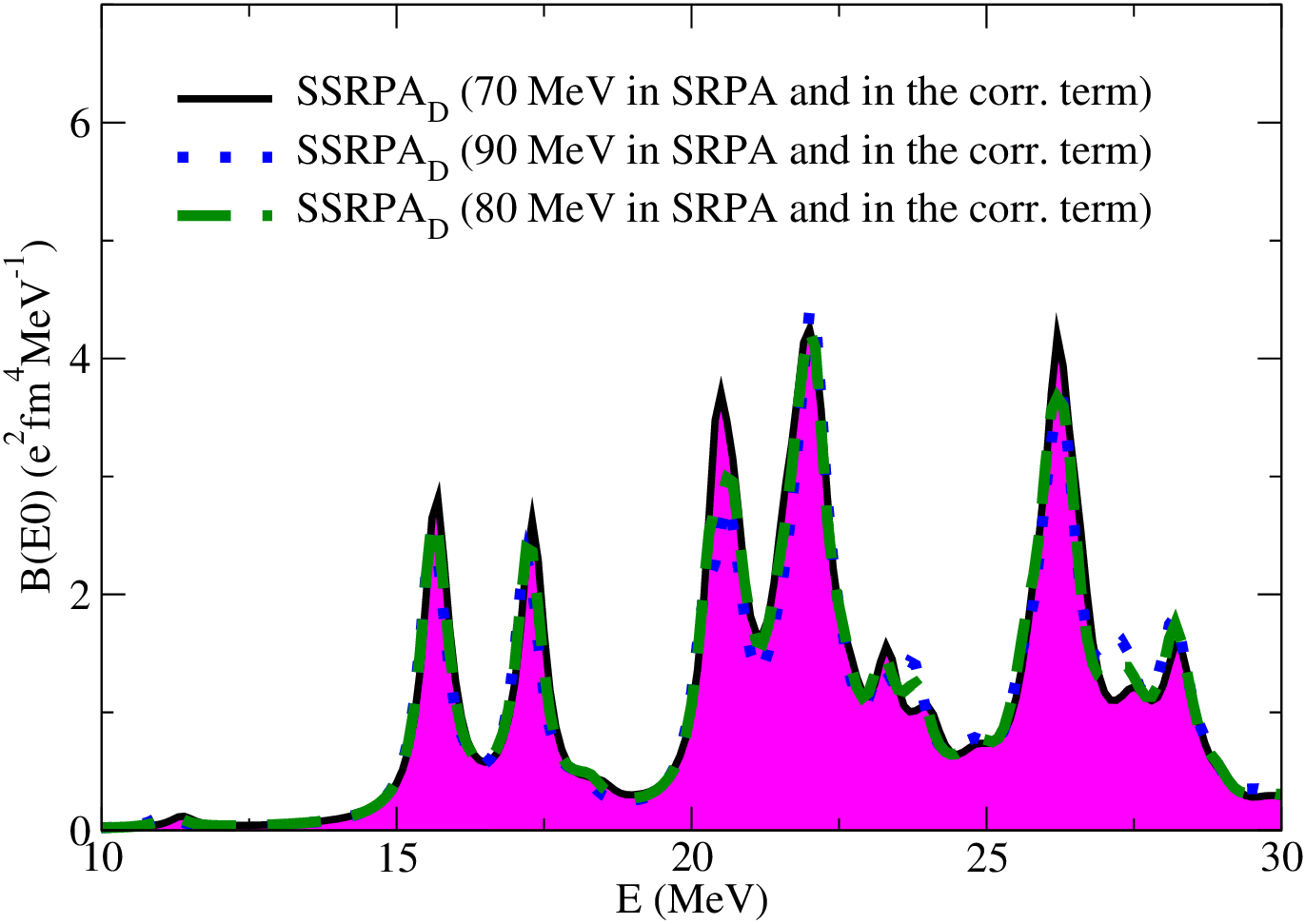}
\caption{(Color online) Isoscalar monopole response in the diagonal
approximation with cutoff for the correction terms at 70 (black line and
magenta area), 80 (green dashed line), and 90 (blue dotted line) MeV. }
\label{checkgiant}
\end{figure}

Before comparing our results with those obtained in the RPA and
with the experimental values, we analyze the robustness of the
subtracted SRPA model.  We focus on the $0^+$ channel.  We initially carry out
standard SRPA calculations with several cutoffs up to 90 MeV on the $2p2h$
configurations, finding, as expected, that the shift to lower energies (with
respect to the RPA spectrum) becomes more and more pronounced as the cutoff
energy increases \cite{gamba1}. The number of $2p2h$ configurations is too
large in these high--cutoff SRPA calculations to invert the matrix $A_{22'}$,
and so when examining the subtraction correction we use the diagonal
approximation.  Fig.\ \ref{checkgiant} shows the resulting isoscalar monopole
responses, with cutoffs for the correction terms at 70, 80, and 90 MeV.  In
each case, this cutoff is the same as that in the corresponding unsubtracted
SRPA calculation. The three strength functions are very similar.  The same is
true of the low--lying 0$^+$ state. Its energy with the 70, 80, and 90 MeV
cutoff is 6.26 MeV, 6.13 MeV, and 5.96 MeV, respectively. The difference
between the highest and lowest of these numbers is only 5\%.  We can conclude
that the subtraction procedure not only rectifies the SRPA energy shifts for
giant resonances, but also provides much more robust (cutoff--insensitive)
predictions for both giant resonances and low--lying states. 

%\begin{figure}
%\includegraphics[scale=0.35]{fig8four-danilo}
%\caption{(Color online) Energy of the first 0$^+$ state calculated with the $DCOrr$ scheme, with a cutoff of 70, 80, and 90 MeV for the corrective term. Each $DCorr$ calculation is done from a SRPA calculation with the same cutoff on the $2p2h$ configurations.}
%\label{checklow}
%\end{figure}

\section{Comparison with RPA and experiment}

We turn finally to the quality of the subtracted SRPA results in comparison
with those of the ordinary RPA and with 
experiment.  We again restrict ourselves to the monopole and quadruople cases.
For these comparisons, we carry out fully the subtraction, with the maximal
cutoffs given earlier (50 and 70 MeV for the monopole and quadrupole channels,
respectively).
\begin{figure}
\includegraphics[scale=0.35]{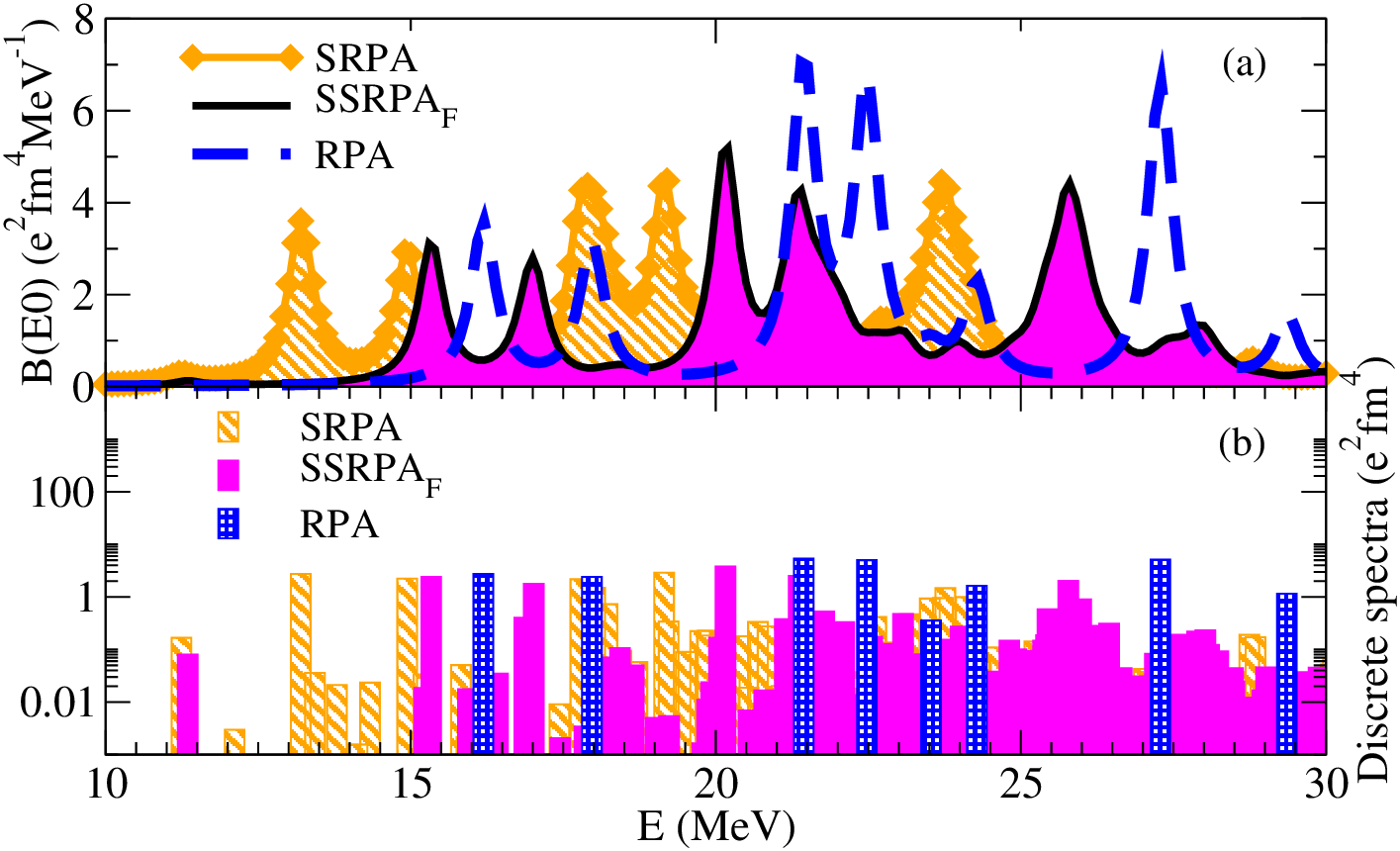}
\caption{(Color online) (a): Isoscalar monopole response in the standard SRPA
(orange diamonds and orange area), RPA (blue dashed line), and
the SSRPA$_F$ (black solid line and magenta area); (b): Discrete
spectra (binned strength) obtained with the SRPA (orange dashed
bars), the RPA (blue dotted bars) and the
SSRPA$_F$ (magenta solid bars).  }
\label{expmono}
\end{figure}

\begin{figure}
\includegraphics[scale=0.35]{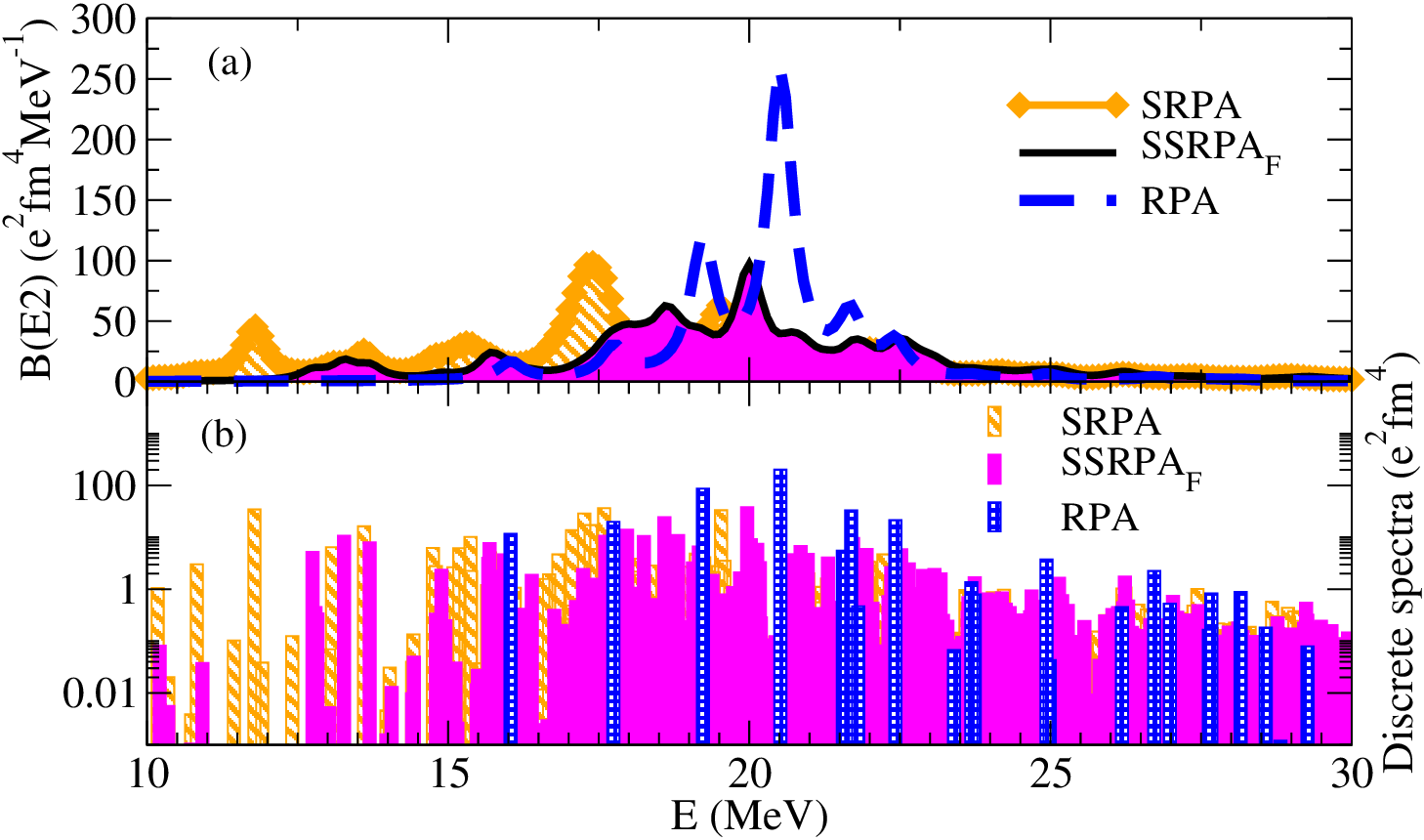}
\caption{(Color online) Same as in Fig.\ \ref{expmono} but for the quadrupole
strength. }
\label{expquadru}
\end{figure}

The top panels of Figs.\ \ref{expmono} and \ref{expquadru}, respectively,
compare the monopole and quadrupole strength distributions of the ordinary
SRPA, the SSRPA$_F$, and the RPA.  We observe that the strong shift that the
ordinary SRPA provides with respect to the RPA is significantly
reduced by the subtraction procedure.  The lower panels of the same figures
display the discrete (binned) strengths.  The fragmentation and the width of
the excitation are provided by both SRPA models (with and without
subtraction).  They are described in a natural way by the extremely dense
discrete $2p2h$ configurations obtained in both cases. The subtraction
procedure does not affect this feature but only shifts states to higher
energies. The ordinary RPA, by contrast, produces a scattered set of states
without any fragmentation.

\begin{figure}
\includegraphics[scale=0.35]{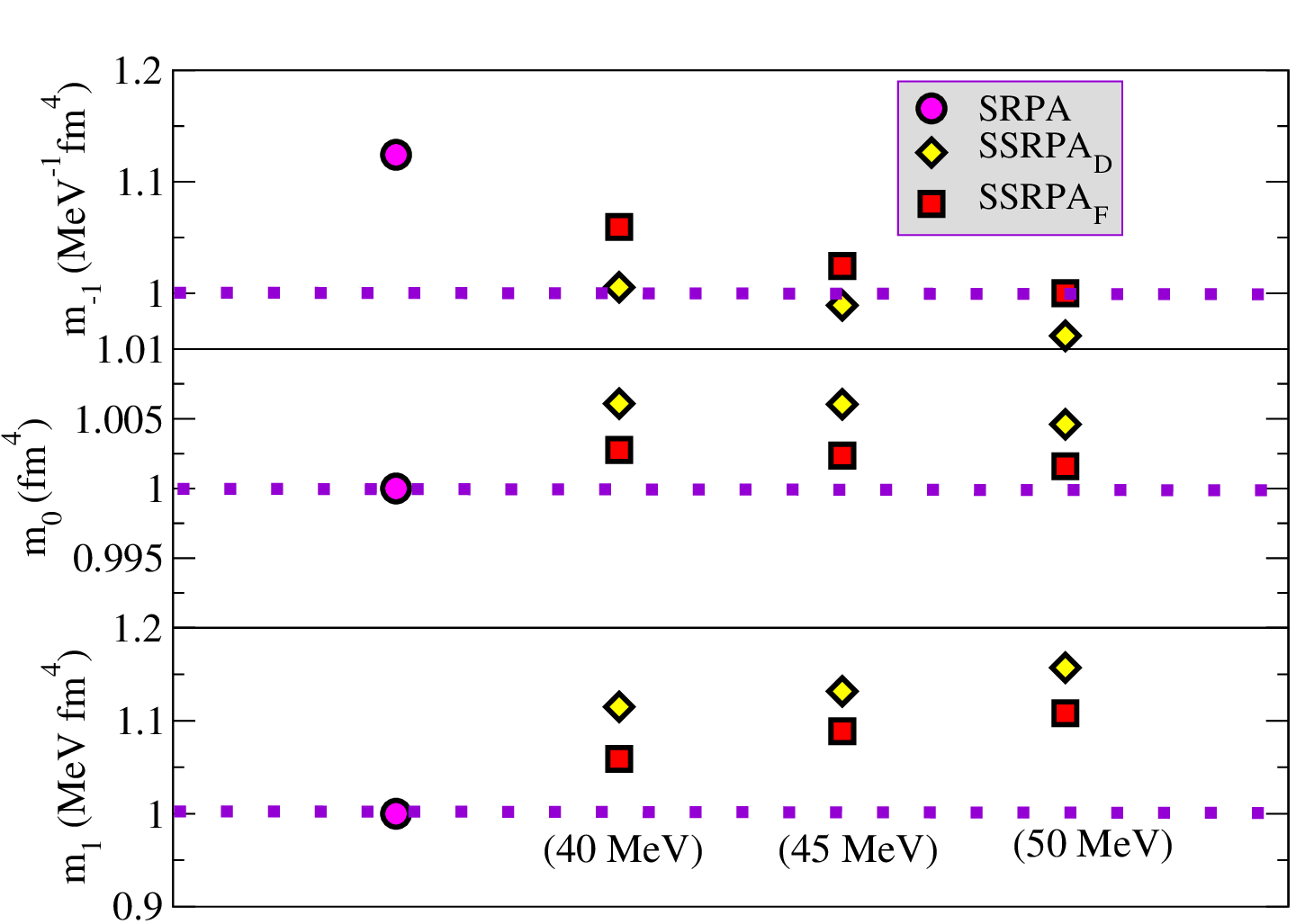}
\caption{(Color online) Ratios of the moments $m_{-1}$, $m_0$, and $m_1$ of the
quadrupole strength distribution in the SRPA (purple circles), the 
SSRPA$_F$ (red squares), and the SSRPA$_D$ (yellow diamonds)
to those in the RPA for increasingly high cutoffs in the correction terms, at 40, 45, and 50 MeV.}
\label{mom}
\end{figure}

Before comparing with experiment, we analyze the moments $m_{-1}$, $m_0$, and
$m_1$ of the strength--distribution, for the isoscalar quadrupole case, as an
illustration.  In Fig.\ \ref{mom} we show the ratios of
these moments, calculated in the SRPA,
the SSRPA$_F$, and the SSRPA$_D$, to those of the
RPA.  We have varied the cutoff energy in the correction terms for the
subtracted calculations. One observes first that $m_0$ and $m_1$ are the same
in the RPA and in the SRPA, as expected.  The corresponding subtracted--SRPA
moments are different, however.  The upward shift from subtraction means that
$m_1$ must be larger with subtraction than without and $m_{-1}$, as noted
earlier, must be smaller; the figure bears these conclusions out.
Unsurprisingly, in addition, the full subtraction produces (at maximal cutoff)
values that are closer to the RPA than does the diagonal approximation.  The
inverse moment $m_{-1}$ is exactly the same in the SSRPA$_F$ with maximal
cutoff and in the RPA, as it must be (see Sec. III). The equality holds only
when the calculation is fully coherent (full inversion and same $2p2h$ space in
the matrices and in the correction term).  

\begin{figure}
\includegraphics[scale=0.35]{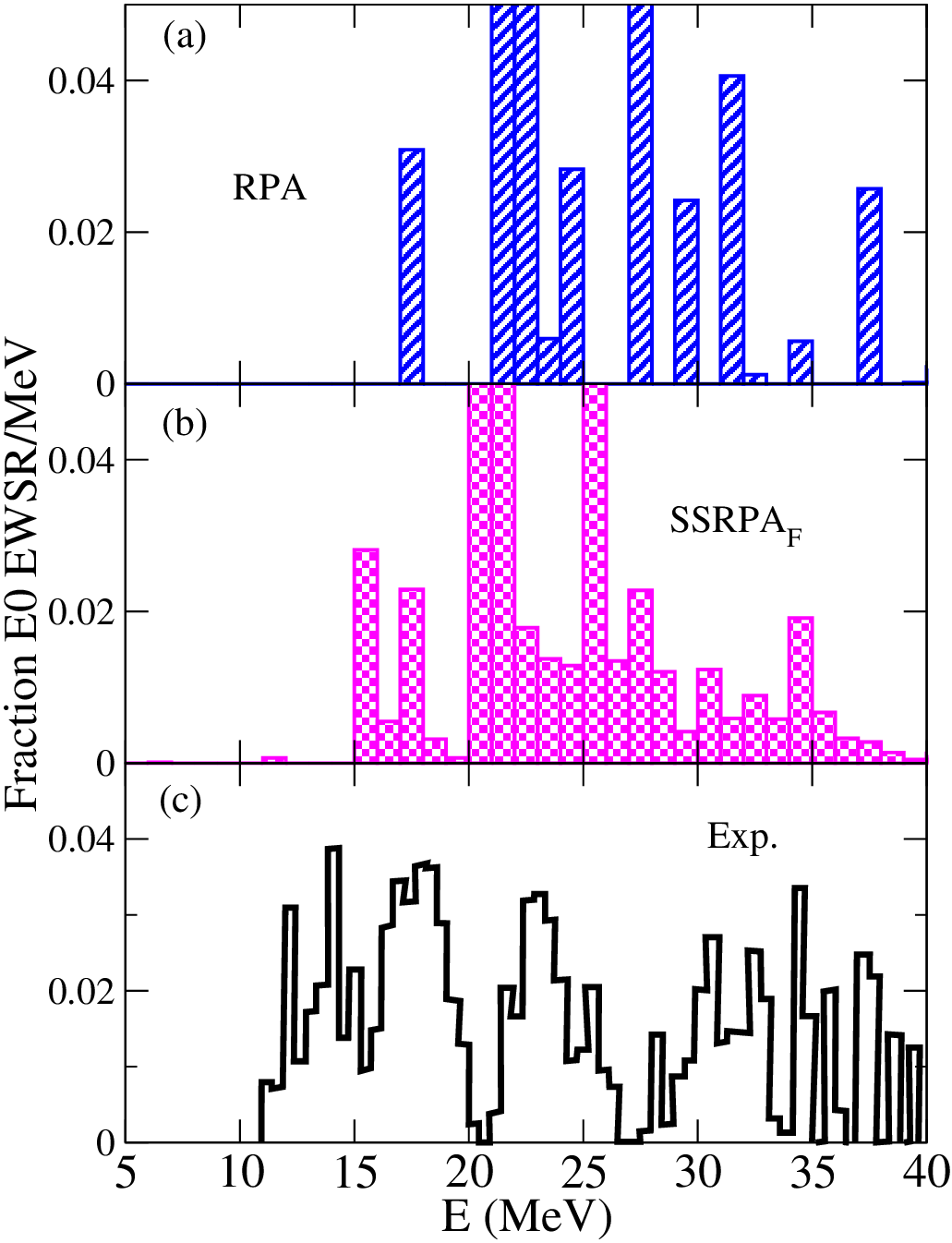}
\caption{(Color online) Fraction of the $E0$ EWSR in the RPA (a) and in
the SSRPA$_F$ (b). The experimental fraction, from Ref.\
\cite{expemonoquadru} is in panel (c). }
\label{ex1}
\end{figure}

\begin{figure}
\includegraphics[scale=0.35]{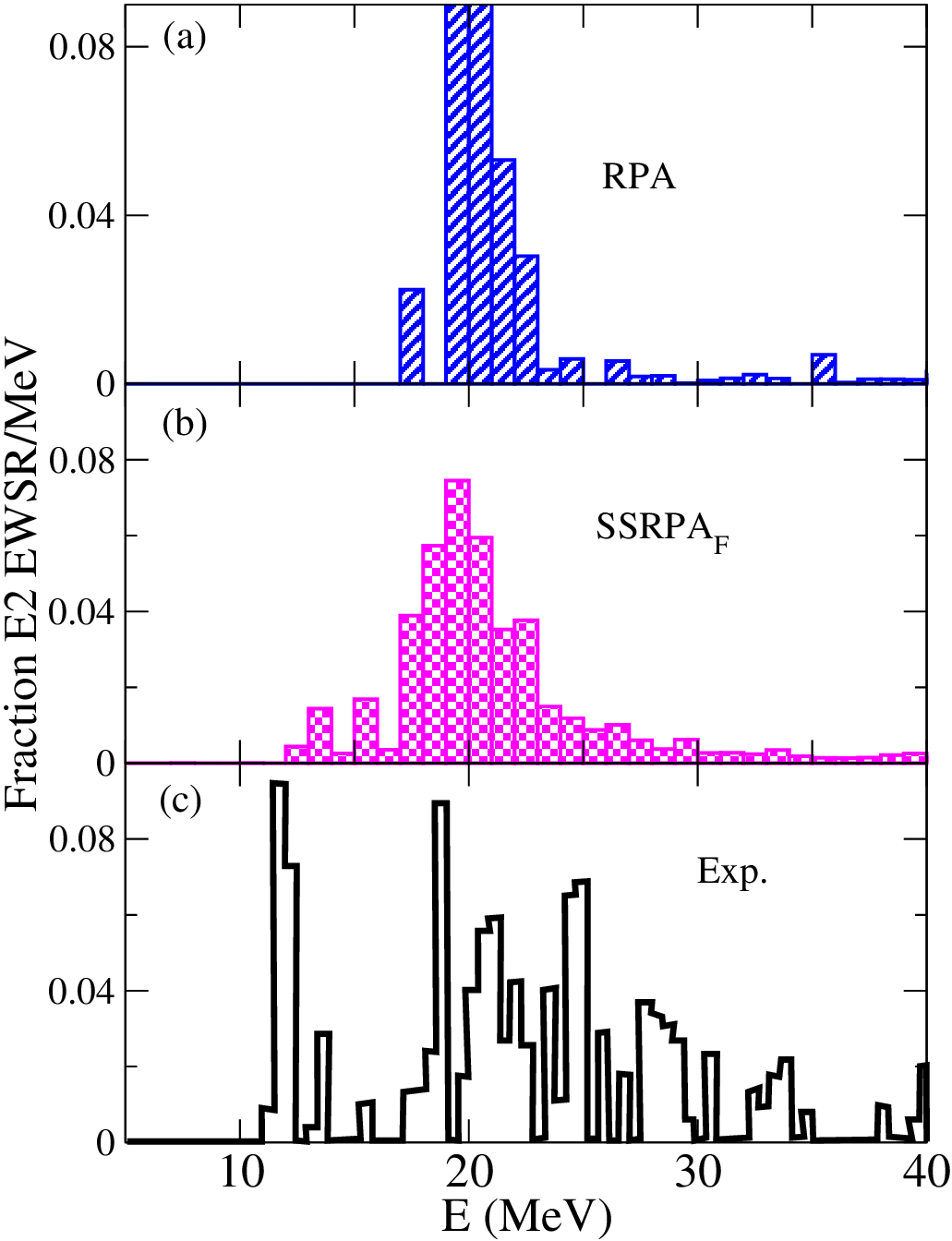}
\caption{(Color online) Same as in Fig.\ \ref{ex1} but for the $E2$ EWSR. }
\label{ex2}
\end{figure}

Finally, we compare our results with experiment.  Figs.\ \ref{ex1} and
\ref{ex2} compare the RPA and subtracted SRPA spectra with experimental
strength distributions \cite{expemonoquadru}.  We have extracted the
experimental fractions of the EWSR
from the top and the bottom panels of Fig.\ 6 in Ref.\  \cite{expemonoquadru}.
To compare with our results we have multiplied our fractions of the monopole
EWSR by 0.48 and the quadrupole EWSR by 0.53.  The reason is that only the 48\%
of the $E0$ EWSR and the 53\% of the $E2$ EWSR have been measured
\cite{expemonoquadru}. Note that the scales of the vertical axes in panels (a)
and (b) are chosen equal to those of the corresponding panels (c) for an easier
comparison.

We observe that, for the monopole case, 
the RPA produces a strength distribution
that is higher in energy than experiment by several MeV.  This feature of the
RPA for the nucleus $^{16}$O is well known.  In Ref.\ \cite{expemonoquadru}, for example,
the RPA results of Ref.\ \cite{ma} were shifted downwards by 4.2 MeV to agree
with the measured centroid.  The SRPA results produced here (without
subtraction) would be in much better agreement with experiment because there is
much more strength at low energies, and in particular there is non--negligible
strength between 10 and 15 MeV (see Fig.\  \ref{expmono}).  One should keep in
mind, however, that the SRPA results without subtraction are strongly cutoff
dependent; strength would be shifted to even lower energies if the cutoff
energy were increased significantly, worsening the agreement with experiment.
In contrast, as shown in this work, the results obtained with the
subtraction procedure are quite stable against variations in the cutoff.
Though the subtracted SRPA produces a 
 reduced downward shift from the RPA than
the ordinary SRPA with the cutoff chosen here (and as a result appears
to agree less well with experiment) the shift is nevertheless in the right
direction: Some strength appears between 10 and 15 MeV and there is much more
strength between 15 and 20 MeV than in the RPA.  

For the quadrupole case, the shift between the results of the RPA and
those of the subtracted SRPA is less pronounced. 
The main difference between the two response functions is
in the fragmentation, which is obviously much better described 
within the subtracted SRPA. 
  Some strength is found between 10 and 15 MeV, as in the experimental
distribution, while in the RPA the strength starts at higher energy. 

\begin{figure}[b]
\includegraphics[scale=0.35]{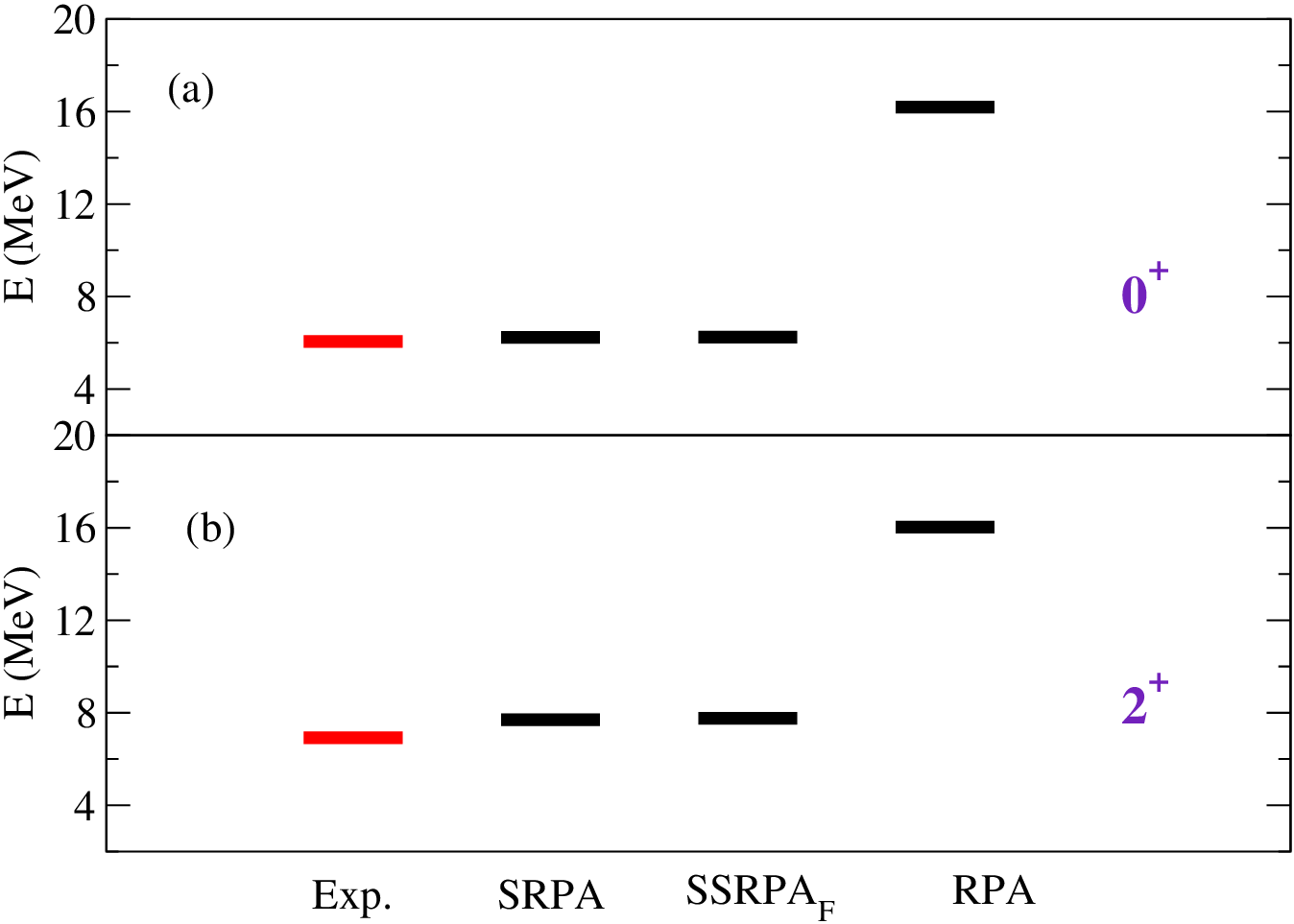}
\caption{Comparison of values from the standard SRPA, the SSRPA$_F$, the RPA,
and experiment for the energy of the first low--lying 0$^+$ (a) and 2$^+$ (b)
states. }
\label{explow}
\end{figure}

Let us now analyze the low--lying states.
Fig.\ \ref{explow} compares the SRPA, SSRPA$_F$, and RPA energies with those
from experiment \cite{low} for the first 0$^+$ and 2$^+$ states.  The RPA
values are too high in energy and the SRPA results (with or without
subtraction) are in good agreement with experiment.  As already noted, the
subtraction does not modify the energies of such states because their most
important configurations are $2p2h$. 
 
\section{Conclusions}

We have applied to the SRPA a subtraction procedure proposed by Tselyaev
\cite{tsela2007} some years ago to overcome problems related to double counting
in certain beyond--mean--field calculations.  Ref.\ \cite{tsela2013} showed
that the subtraction method in extended RPA models, such as the SRPA, leads to
stable solutions (the Thouless theorem may be extended and the stability
condition satisfied). 

We have presented applications to the nucleus $^{16}$O with the Skyrme
interaction SGII. We have verified that the subtracted SRPA provides very
robust predictions, which are stable and very weakly cutoff dependent.
Furthermore, the fulfillment of the stability condition, together with the
elimination of double counting, substantially reduces the large anomalous shift
downwards that the ordinary SRPA systematically produces with respect to the
RPA strength. 

This implementation of the SRPA model opens the door to numerous applications
to reliably assess the effects of multiparticle--multihole configurations on
the excitation of medium--mass and heavy nuclei. 

\begin{acknowledgments}
M.G.\ acknowledges the support of the International Associated 
Laboratory COLL--AGAIN. 

J.E.\ acknowledges the support of the U.S.\ Department of
Energy through Contract No.\ DE-FG02-97ER41019.
\end{acknowledgments}

%\begin{widetext}
%\begin{center}
%\begin{table}[htbp]
%\centering
%      \begin{tabular}{cccccccccc}\hline\aline
%                 $t_0$ (MeV fm$^{3}$)  &$t_1$(MeV fm$^{5}$)& $t_2$(MeV fm$^{5}$)& $T_3$(Mev fm$^5$)&$x_0$ &$x_1$&$x_2$&$x_3$ & $\alpha$ \%\\hline
% -1340.089&  1343.713& -1826.053&    3154.567&0.000458&  6.469&  4.328&  3.109&2/3 \\\hline\hline
%      \end{tabular}
%      \caption{Parameter set obtained in the fit.}
%\label{fit_skpfinite}
%\end{table}
%\end{center}
%\end{widetext}

%\begin{figure}
%\includegraphics[scale=0.3]{fit0}
%\caption{(Color online) Refitted EoS (dashed line) compared with the SLy5--mean--field EoS (solid line).}
%\end{figure}

%
%
%
%-----------------------------------


\begin{thebibliography}{10}
\bibitem{providencia} J. da Providencia, Nucl. Phys. 61, 87 (1965).
\bibitem{Yannouleas} C. Yannouleas, Phys. Rev. C 35, 1159 (1987). 
\bibitem{papa1} P. Papakonstantinou and R. Roth, Phys. Lett. B 671, 356 (2009).
\bibitem{papa2} P. Papakonstantinou and R. Roth, Phys. Rev. C 81, 024317 (2010).
\bibitem{gamba1} D. Gambacurta, M. Grasso, and F. Catara, Phys. Rev. C 81, 054312 (2010).
\bibitem{gamba2} D. Gambacurta, M. Grasso, and F. Catara, J. Phys. G 38, 035103 (2011).
\bibitem{gamba3} D. Gambacurta, M. Grasso, and F. Catara, Phys. Rev. C 84, 034301 (2011). 
\bibitem{gamba4} D. Gambacurta, M. Grasso, V. De Donno, G. Co’, and F. Catara, Phys. Rev. C 86, 021304(R) (2012). 
\bibitem{skyrme1} T.H.R. Skyrme, Philos. Mag. 1, 1043 (1956).
\bibitem{skyrme2} T.H.R. Skyrme, Nucl. Phys. 9, 615 (1959). 
\bibitem{vautherin} D. Vautherin and D. Brink, Phys. Rev. C 5, 626 (1972). 
\bibitem{rifeGogny1} J. Decharg\'e and D. Gogny, Phys. Rev. C 21, 1568 (1980). 
\bibitem{rifeGogny2} J.F. Berger, M. Girod, D. Gogny, Comput. Phys. Commun. 63, 365 (1991). 
\bibitem{moghrabi} K. Moghrabi, M. Grasso, G. Col\`o, and N. V. Giai, Phys. Rev. Lett. 105, 262501 (2010).
\bibitem{tsela2007} V.I. Tselyaev, Phys. Rev. C 75, 024306 (2007). 
\bibitem{nonrela1} E.V. Litvinova and V.I. Tselyaev, Phys. Rev. C 75, 054318 (2007).
\bibitem{nonrela2} V. Tselyaev, {\it et al.}, Phys. Rev. C 79, 034309 (2009).
\bibitem{nonrela3} A. Avdeenkov, S. Goriely, S. Kamerdzhiev, and S. Krewald, Phys. Rev. C 83, 
064316 (2011).
\bibitem{rela1} E. Litvinova, P. Ring, and V. Tselyaev, Phys. Rev. C 75, 064308 (2007).
\bibitem{rela2} E. Litvinova, P. Ring, and V. Tselyaev, Phys. Rev. C 78, 014312 (2008).
\bibitem{rela3} E. Litvinova, P. Ring, and V. Tselyaev, Phys. Rev. Lett. 105, 022502 (2010).
\bibitem{tsela2013} V.I. Tselyaev, Phys. Rev. C 88, 054301 (2013).
\bibitem{thouless} D.J. Thouless, Nucl. Phys. 22, 305 (1961).
\bibitem{papa2014} P. Papakonstantinou, Phys. Rev. C 90, 024305 (2014).
\bibitem{ringshuck} P. Ring and P. Shuck, {\it The Nuclear Many-Body Problem} (Springer-Verlag, Berlin, 1980). 
\bibitem{gambacatara} D. Gambacurta and F. Catara, Phys. Rev. B 81, 085418 (2010).
\bibitem{sgii} N.V. Giai, and H. Sagawa, Phys. Lett. B 106, 379 (1981); Nucl. Phys. A 371, 1 (1981).
%\bibitem{schirmer} J. Schirmer and G. Angonoa, J. Chem. Phys. 91, 1754 (1989).
\bibitem{dft} see, e.g., \textit{A Primer in Density Functional Thery}, C.
Fielhaus F. Ngueira, and M. Marques, eds., Springer (1997). 
\bibitem{hohenbergkohn} P. Hohenberg and W. Kohn, Phys.\ Rev.\ 136, 8864 (1964).
\bibitem{kohnsham} W. Kohn and L.J. Sham, Phys.\ Rev.\ 140, A1133 (1965).
\bibitem{engel} J. Engel, Phys.\ Rev.\ C 75 014306 (2007). 
\bibitem{rungegross} E. Runge and E.K.U. Gross, Phys.\ Rev.\ Lett. 52, 997
(1984).
\bibitem{expemonoquadru} Y.-W. Lui, H.L. Clark, and D.H. Youngblood, Phys. Rev. C 64, 064308 (2001).
\bibitem{ma} Z. Ma, N. Van Giai, H. Toki, and M. L'Huillier, Phys. Rev. C 55, 2385 (1997). 
\bibitem{low} F. Ajzenberg-Selove, Nucl. Phys. A 375, 1 (1985). 


\end{thebibliography}
\end{document}